\documentclass[12pt]{iopart}

\usepackage{graphicx} 
\usepackage{hyperref}
\usepackage{wasysym} 

\begin{document}


\title[Short title]{A cryogenic test-mass suspension with flexures operating in compression for third-generation gravitational-wave detectors}

\author{
Fabi\'an E.~Pe\~na Arellano$^{1,2,3}$,
Nelson L.~Leon$^{3}$,
Leonardo Gonz\'alez L\'opez$^{1}$,
Riccardo DeSalvo$^{3,4,5}$,
Harry Themann$^{6}$,
Esra Zerina Appavuravther$^{7}$,
Guerino Avallone$^{8}$,
Francesca Badaracco$^{9}$,
Mark A. Barton$^{10}$,
Alessandro Bertolini$^{11}$,
Christian Chavez$^{3}$,
Andy Damas$^{3}$,
Richard Damas$^{3}$,
Britney Gallego$^{3}$,
Eric Hennes$^{11}$,
Gerardo Iannone$^{5}$,
Seth Linker$^{6}$,
Marina Mondin$^{3}$,
Claudia Moreno$^{1}$,
Kevin Pang$^{3}$,
Stefano Selleri$^{12}$,
Mynor Soto$^{3}$,
Flavio Travasso$^{13,14}$,
Joris Van-Heijningen$^{11,15}$,
Fernando Velez$^{3}$,
Morgane Zeoli$^{16,17}$.
}

\address{$^{1}$University of Guadalajara, University Centre for Exact Sciences and Engineering (CUCEI), Blvd.~Gral.~Marcelino Garc\'ia Barrag\'an 1421, Col.~Ol\'impica, 44430 Guadalajara, Jalisco, M\'exico.}
\address{$^{2}$Institute for Cosmic Ray Research, KAGRA Observatory, The University of Tokyo, 238 Higashi-Mozumi, Kamioka-cho, Hida City, Gifu 506-1205, Japan.}
\address{$^{3}$California State University, Los Angeles, College of Engineering, Computer Science and Technology, 5151 State University Dr, Los Angeles, CA 90032, USA.}
\address{$^{4}$University of Sannio, Department of Engineering , Corso Garibaldi 107, I-82100 Benevento, Italy.}
\address{$^{5}$National Institute for Nuclear Physics (INFN), Gruppo Collegato di Salerno, Sezione di Napoli, via Giovanni Paolo II, Fisciano (SA), Italy.}
\address{$^{6}$California State University, Los Angeles, Department of Physics and Astronomy, 5151 State University Dr, Los Angeles, CA 90032, USA.}
\address{$^{7}$Max Planck Institute for Gravitational Waves, Albert Einstein Institute -- GEO 600, Callinstra{\ss}e 38, 30167 Hannover, Germany.}
\address{$^{8}$Department of Physics “E. R. Caianiello”, University of Salerno, Via Giovanni Paolo II 132, Fisciano (SA), I-84084, Italy.}
\address{$^{9}$University of Genova, Department of Physics, via Dodecaneso 33, 16146 Genova, Italy.}
\address{$^{10}$Institute for Gravitational Research, University of Glasgow, Glasgow G12 8QQ, United Kingdom.}
\address{$^{11}$Nikhef, Science Park 105, 1098 XG Amsterdam, Netherlands.}
\address{$^{12}$University of Florence, Dept.~of Information Engineering; Via di S.~Marta, 3, 50139 Firenze, Italy.}
\address{$^{13}$University of Camerino, School of Science and Technology, Physics Division, Via Madonna delle Carceri, 9B, 62032 Camerino, MC, Italy.}
\address{$^{14}$National Institute for Nuclear Physics (INFN), Sezione di Perugia - via Pascoli, 06125 Perugia, Italy.}
\address{$^{15}$Vrije Universiteit Amsterdam, Department of Physics and Astronomy, De Boelelaan 1100, 1081 HZ Amsterdam, Netherlands.}
\address{$^{16}$Centre for Cosmology, Particle Physics and Phenomenology (CP3), UCLouvain, Chemin du cyclotron 2, B-1348 Louvain-la-Neuve, Belgium.}
\address{$^{17}$Precision Mechatronics Laboratory, A\&M Dept, ULi\`ege, All\'ee de la D\'ecouverte 9, B-4000 Li\`ege, Belgium.}

\ead{naibaf.omsare@gmail.com}
\vspace{10pt}
\begin{indented}
\item[]January 2025
\end{indented}

\begin{abstract}

This paper presents an analysis of the conceptual design of a novel silicon suspension for the cryogenic test-mass mirrors of the low-frequency detector of the Einstein Telescope gravitational-wave observatory. In traditional suspensions, tensional stress is a severe limitation for achieving low thermal noise, safer mechanical margins and high thermal conductance simultaneously. 
In order to keep the tensional stress sufficiently low, we propose the use of rigid beams with large cross sections, combined with short flexures under compressional load. This configuration takes advantage of the many times higher strength of silicon in compression to respect to its strength in tension.
 The flexures are mechanically robust and at the same time soft in the working direction, thus producing low suspension thermal noise and, by being short, provide high thermal conductance for cryogenic cooling.
 The rigid beams, located between the test mass and an intermediate mass, allow the elimination of the recoil mass used conventionally for applying control forces for interferometer lock, and the use of optical anti-springs to reduce the pendulum resonant frequency to further improve the vibration isolation of the test mass. The configuration has the capability to reach a lower mirror operational temperature, which is expected to produce a substantial reduction of the thermal noise in the mirrors of the interferometer.

\end{abstract}

%
%
%
%
%

%
%


\section{ \label{sec:introduction} Introduction}

The Einstein Telescope (ET) is a project for the construction of a third-generation interferometric gravitational-wave detector. It is being designed to extend the astronomical observation frequency down to a few hertz, to detect inspirals of heavier, intermediate-mass binary back hole coalescences up to cosmological distances, and also to provide early warning for multi-messenger events \cite{Maggiore-2020,ET-feasibility}.

One of the most difficult requirements is to minimize thermal noise. Because of quantum noise restrictions, the ET will employ two kinds of detectors, a cryogenic one for low-frequency detection and a room temperature one for the high-frequency end. The design configuration proposed here is intended for the cryogenic detector.

Thermal noise of the test-mass mirrors and of their suspensions is a fundamental limit to the detector low-frequency sensitivity.
Thermal noise refers to the fluctuations of the mirror positions arising  from mechanical dissipation at finite temperatures, as described by the fluctuation-dissipation theorem and Levin's theorem \cite{PhysRev.83.34,Kubo-1966,PhysRevD.57.659}. The two strategies considered effective to reduce it are to operate at cryogenic temperatures and to minimize the amount of energy being dissipated in each oscillation.
In turn, there are two strategies to reduce the amount of energy that can be dissipated in an oscillating pendulum. The first is to build the suspension with a material that exhibits low intrinsic structural dissipation. The second is to choose a mechanical configuration that minimizes the amount of elastic potential energy, which is the only one exposed to the dissipation process.
This requires making the pendulum elastic constant negligible with respect to the gravitational spring constant, using a longer pendulum, or both.

The light power stored in the Fabry-Perot cavities  is kept low to minimize the radiation pressure noise on the mirrors, but an optical power of 18\,kW is still needed to achieve the desired sensitivity \cite{ET-2020-report}. A tiny fraction of this  power is dissipated into heat in the reflective coating of the mirrors and in the substrates of the input test masses. It is anticipated that up to 0.5\,W will be dissipated in each of the 200\,kg mirrors \cite{ET-2020-report}.
Thermal radiation is negligible at cryogenic temperatures; therefore, the system must be able to extract sufficient heat via thermal conduction to maintain a stable low operational temperature of the suspensions and mirrors. Since the suspension is the only way in which the cryogenic mirror couples mechanically to the external environment, a pendulum is also required to be an effective means for extracting heat from it. 

A typical structure optimized to extract heat will have a large cross section, a short length, or both \cite[eq.~18-32]{halliday2013fundamentals}, making the pendulum rigid and thus increasing the elastic energy stored in the material during oscillations and the suspension thermal noise.

Monocrystalline silicon and sapphire have been identified as the best materials for this application due to their low mechanical dissipation and high thermal conductivity at cryogenic temperatures, with silicon being preferred because it is softer with a relatively small Young’s modulus. However, the limited strength of all monocrystalline materials in tension  imposes the additional constraint of requiring large cross sections. When subjected to even limited tensional stresses, cleavage along crystalline planes leads to catastrophic failure, with microscopic surface defects being sufficient to initiate the process. Therefore, it is required that crystalline components under tensional or shear stress have sufficiently large cross sections to drastically reduce the tensional stress.

This document describes a method to solve this conundrum by taking advantage of the many times higher compressive strength of silicon with respect to its tensile strength \cite{azo-materials-si}. The key elements are flexures that work in compression \cite[p.~70]{Blom-PhD-2015}, and therefore, without the risk of cleavage and capable of withstanding higher stress. These flexures can be thin and, therefore flexible, but also short to allow efficient heat extraction.

The values of the strength of silicon used in this paper,  165 and 3200\,MPa for tension and compression respectively, are provided by a commercial vendor and are rather conservative \cite{azo-materials-si}, for they likely correspond to unpolished samples \cite{azo-materials-si}. In recent months, within the ET Collaboration, innovative  techniques of manufacturing \cite{VIEIRA2024127549} and polishing have been developed, increasing the strength of silicon rods in tension. Some gain may be expected for compressive strength as well, because surface defects may also trigger shear failure along inclined crystalline planes, the typical failure mode of elements in compression. Whereas the details of the components described below will need to be reviewed as more definitive data and criteria become available, the goal of the paper of presenting the feasibility of a concept rather than a final design stands.

Therefore, this study is limited to a conceptual design. It describes qualitatively how to use flexures to design a double-pendulum cryogenic payload and provides a semi-quantitative evaluation of a pendulum hinged with flexures. The aim is to point out the technical opportunities offered for future developments. Considering this scope, the selected dimensions of the components are reasonable, nevertheless, a true optimization can only be performed by considering all the requirements from a holistic point of view. 

In this article, the term {\em suspension} refers to a multiple-pendulum cryogenic assembly, which is the payload hanging from a room-temperature  multi-stage vibration isolation chain.

While guided by the necessity to generate a soft system with a high thermal conductance, the ensuing design offers solutions that may apply also to room-temperature detectors. Between these perhaps the most interesting is the opportunity to eliminate the recoil mass used to control the mirror during interferometer lock acquisition.

Section \ref{sec:concept} describes the hardware configuration and the conceptual motivation of its distinctive features. Sec.~\ref{sec:flexure} reports on the flexure design and its mechanical performance. Sec.~\ref{sec:beam} describes the suspension beam features, how to mitigate the resonances it introduces, the centre of percussion effect and the possible use of an optical anti-spring to reduce the pendulum resonant frequency. Sec.~\ref{sec:spring-blades} reports on the design of vertical blade springs. Sec.~\ref{sec:temperature} reports on the mirror cryogenic operating temperature and the cooling time from room temperature. Finally, Sec.~\ref{sec:thermal-noise} presents the displacement thermal noise produced by the suspension.  

\section{\label{sec:concept} Concept of the proposed solution}

This section illustrates an innovative mechanical configuration that uses flexures in compression in a cryogenic suspension and explains the motivations for its most important features.

\begin{figure}
	\centering
	\includegraphics{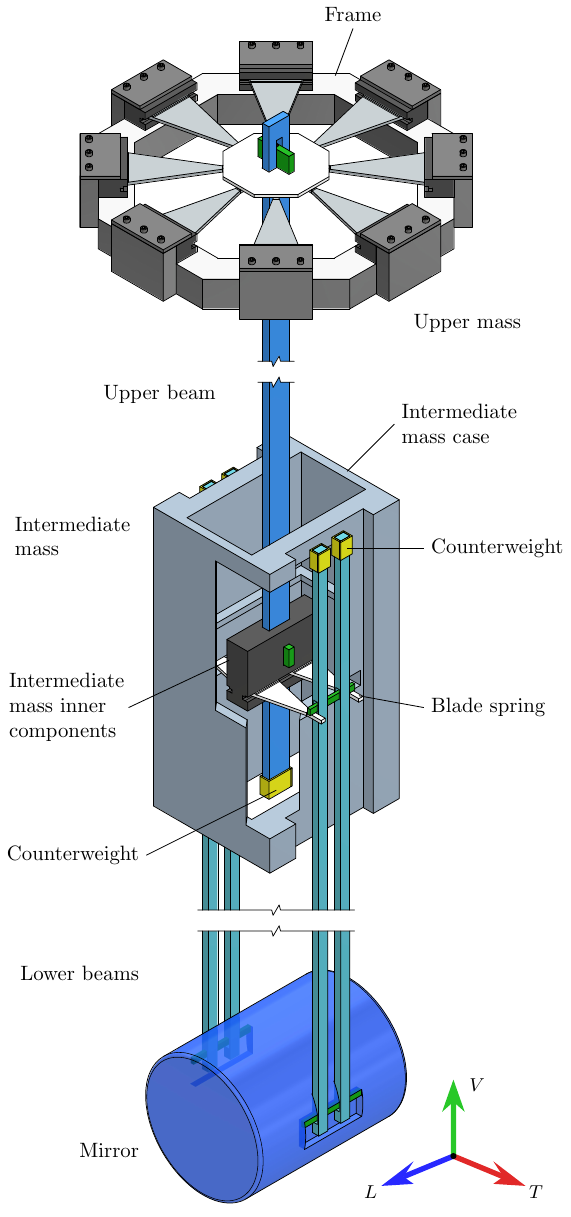}	
	\caption{ \label{fig:payload} View of the suspension assembly, where the lower and upper beams are represented shorter and a section of the outer case of the intermediate mass has been cut away to show the inner components. With the exception of the intermediate mass case and the frame of the upper mass, which are made of aluminium, all components are made of silicon. The dimensions of the mirror were taken from Ref.~\cite{ET-materials-database}. The labels $L$, $V$ and $T$ stand for the longitudinal, vertical and transverse directions.} 
 \end{figure}
 
 Figure \ref{fig:payload} shows the complete suspension system. 
 It comprises three main masses: the upper mass, intermediate mass and mirror. The upper mass is intended to hang from a room-temperature seismic attenuation chain via a suspension rod (not shown) and is treated as immobile in this analysis. The lower masses form a double pendulum connected by two levels of suspension beams of rectangular cross section and vertical blade springs. 
A security structure, surrounding all major components and attached to the ceiling of the cryostat, is foreseen but not yet developed. Its aim is to catch the components in case of breakage  and to prevent large excursions from equilibrium positions. 

The double pendulum configuration is intended to provide mechanical isolation from the  thermal noise produced by the seismic attenuation chain and from vibrations which may be introduced by heat links connecting the upper mass frame to the cryogenic chillers. From the mirror at the bottom up to the components supported by the upper mass frame, all components along the heat path are made of monocrystalline silicon, while the case of the intermediate mass and the upper mass frame are structural aluminium.
  
\begin{figure}
	\centering
	\includegraphics{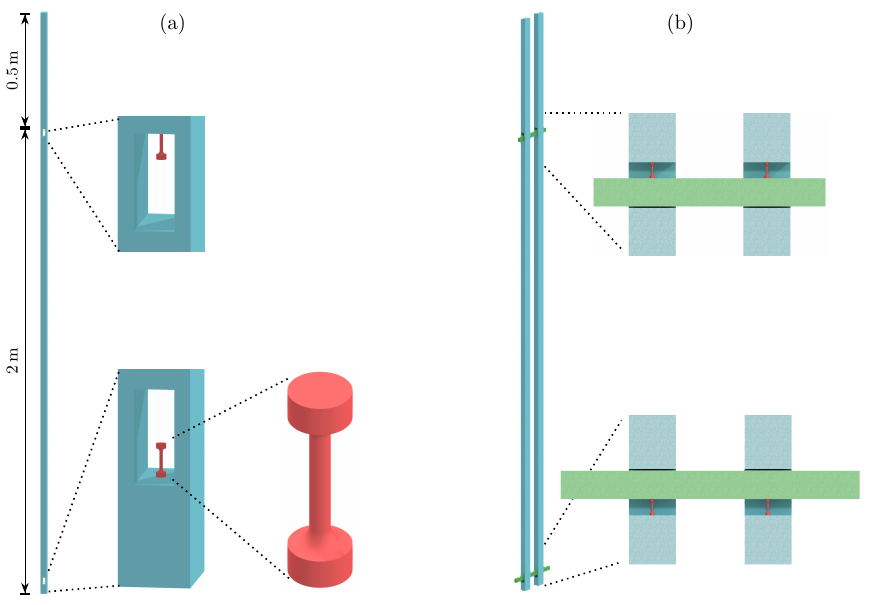}
	\caption{ \label{fig:flexure-and-window} Panel (a) shows one of the suspension beams with detail views of the windows housing the flexures and one of the flexures. Panel (b) shows how two beams are connected with cross beams at the windows. The central cylinder of the flexure is 1.16\,mm in diameter and 6\,mm in length. The length of the pendulum is 2\,m.}
 \end{figure}

Figure \ref{fig:flexure-and-window}(a) shows one of the long suspension beams with two successive detail views to illustrate the flexures, which are very small and hard to see.  The beam is rigid and works in tension, its large cross section maintains the tensional stress well below the breaking stress of the material and allows the length without sacrificing thermal conductance. It can be easily machined from a long silicon boule. A long beam is necessary to achieve a low pendulum resonant frequency.
At its lower end, shown in the detail view, it has a window housing the flexure that supports the mirror. The upper detail view shows a similar window at about four fifths  of the beam length. The beam extends above this window to support a counterweight, whose function will be discussed in Sec.~\ref{sec:beam}.
The flexure, working in compression, is formed by a solid cylinder between two nail heads with fillets at the transition.
Fig.~\ref{fig:flexure-and-window}(b) illustrates how pairs of suspension beams are pre-assembled, connected by two short cross beams. As shown in the detailed cross-section views, a bottom cross beam, shown in green, is supported by the lower flexures while an upper cross beam, also shown in green, supports the assembly at the bottom of the upper flexures. The lower cross beam connects to and supports the mirror while the upper one connects to vertical blade springs.
 
\begin{figure}
	\centering
	\includegraphics{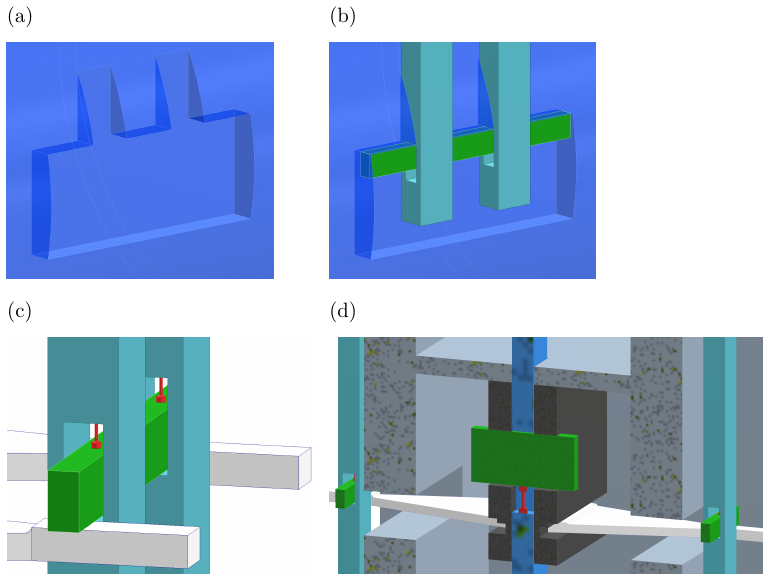}
	\caption{ \label{fig:payload-details} ($a$) Shelves carved at the mirror sides, ($b$) suspension beam structure coupled to the mirror, ($c$) similar structure to couple to intermediate mass blade springs and ($d$) attachment of the blade springs to the intermediate mass.}
 \end{figure}

Figures \ref{fig:payload-details} ($a$) and ($b$) illustrate the coupling of the mirror with the beams. Shelves are carved at the sides of the mirror from which it is directly supported by the cross beams, one of which is shown in green. Fig.~\ref{fig:payload-details}($c$) illustrates how the upper cross beam is supported at the tips of triangular blade springs. Fig.~\ref{fig:payload-details}($d$) shows that the blades are inserted into slots in a central block shown in black and secured in place by the torque produced by the load. At their tips, the blades have thicker straight sections to allow supporting the cross beams horizontally and for applying a pre-stressing force for bending the blade to the correct extent prior to putting  the load in place. A section of the aluminium intermediate mass case is also visible. It is supported by the upper face of the block shown in black.

A single upper suspension beam couples to the central block with a cross beam shown in green and a flexure shown in red. This flexure is larger than the ones described before but otherwise similar to them.
As shown in Fig.~\ref{fig:payload}, the upper beam couples to an upper mass. In an analogous way as it is shown in the upper detail view of Fig.~\ref{fig:flexure-and-window}(b), the beam is supported directly by another large flexure, which is in turn supported by a cross beam fixed on a plate at the centre the upper mass. 
The upper mass is a structural aluminium frame that supports a set of silicon blocks that hold silicon blade springs that work as a vertical mechanical filter. For the calculations reported in this paper, the frame is assumed to be connected by means of soft aluminium wire links to coolers operating at 4\,K. Both the upper and intermediate masses are about 200\,kg, to match the mass of the mirror at the bottom.
 
The advantages and disadvantages and their solutions of using  flexures in compression and rigid suspension beams in tension are listed here, some of which will be further elaborated in the following sections:
\begin{itemize}

	\item The mass of the beams produce a saturation of the attenuation of the vibrations from the stage above. This effect is commonly referred to as the {\em centre of percussion effect} \cite{Akiteru-PhD-2002,Blom-PhD-2015}. The counterweights were designed to mitigate this effect.
	
	\item The upper extensions of the beams can be used as mechanical levers to apply forces on the mirror for interferometer lock control, thus eliminating the requirement of a mirror recoil mass.
	
	\item The beam extension offers the possibility of implementing optical anti-springs to change the pendulum resonant frequency. The anti-spring would be generated by a short, fiber-fed Fabry-Perot mounted between the beam end and the intermediate mass body.
	
	\item The suspension beams introduce  unwanted resonances that must be damped or cooled before starting an observation run. Damping can be passive or active, and reduces the amplitudes down to those corresponding to the thermal bath temperature. Cooling  is active and goes lower than such a temperature. The active system can be implemented with optical sensors and electrostatic actuators \cite{Nelson-master-thesis}. The optical sensor may be the same Fabry-Perot cavity that is later used for the optical anti-springs.
	
	\item The four vertical suspension beams have negligible vertical elastic compliance but they still are subject to unavoidable machining tolerances, so silicon blade springs, triangular in form,  must be used to provide elastic compliance. 
	
	\item As will be reported in Sec.~\ref{sec:spring-blades}, the resonance frequency of these blades is too high to filter out residual noise, mechanical or thermal in origin, from higher stages of seismic isolation. Nevertheless, active anti-springs can be used to provide attenuation. They can be implemented with a similar scheme to the one proposed for beam resonant damping but with a different type of feedback filter \cite{STOCHINO2009737}.	
	
	\item The flexures, being tiny, may be manufactured with $^{28}$Si to take advantage of its higher thermal conductivity, produced by a larger phonon free path as compared to regular silicon \cite{10.1063/1.5017778}. The quality factor of $^{28}$Si has not been measured, but it is reasonable to assume it is higher because it is also limited by the phonon mean free path.
	
	\item The mirror is supported with a purely compressive force applied upwards on horizontal slots carved in its sides. It can be held in the same way during the metrological characterization steps, alternating with corrective polishing steps, needed during fabrication. This is important because each test mass will have a mass of around 200\,kg and its reflecting face will require a predetermined radius of curvature between 5 and 10\,km. With any supporting arrangement different from the one used in the suspension, the  sag of the test mass under its own weight will be different to the one it would have when hanging from it, a condition that would produce a deformation exceeding the curvature tolerances.
	
	\item All the joining surfaces of the assembly are rigorously horizontal and subject to no shear, which allows the use of low-temperature brazing.  Heated and melted under load, any excess of brazing flows out of the contact region, leaving only what is necessary to fill the residual roughness voids, to act as a no-slip adhesive and  to guarantee  thermal contact. Although the contribution of brazing to the thermal noise is expected to be low \cite{PhysRevLett.132.231401}, thickness measurements and calculations are still needed to have a realistic estimate of its contribution. Indium and gallium are convenient because they can be melted in the laboratory or in situ at temperatures of 157$^{\circ}$C and 36$^{\circ}$C respectively. They can be used for different assembly stages, with the highest temperature brazing employed for pre-assembly of the sub-assembly shown in Fig.~\ref{fig:flexure-and-window}(b), and the lower temperature brazing used for the final in-situ assembly, thus dropping the requirement of moving the complete suspension from the outside as in Virgo and KAGRA. This strategy allows longer vertical sizes.
	
	\item Brazed assembly is an important feature for various reasons: mirrors will be extremely expensive and it will be a requirement to be able to assemble and disassemble the suspensions easily, components can be machined separately with higher tolerances in their final dimensions, and the possibility of disassembly allows rearranging misaligned components and also replacing damaged ones.

\end{itemize}

\section{ \label{sec:flexure} Flexure design}

Choosing the dimensions of the flexures is crucial for the suspension performance. A successful design requires an adequate compromise between three conflicting requirements. As mentioned in Sec.~\ref{sec:introduction}, two requirements are low suspension thermal noise and high thermal conductance. The third one is having the unwanted resonances of the suspension as high as possible, outside of the frequency range of the astronomical observations if possible.  
In this section, selection criteria are presented and a choice of flexure dimensions is made. Although this selection is reasonable within the limited scope of this analysis, a final design must take into account the detector requirements in a holistic way.

The performance was calculated approximately with numerical finite-element analysis using both the COMSOL and Ansys software packages. After allowing for the different methods for specifying input parameters in the two programs, good agreement in the output was achieved. For all components, the crystal direction Miller index [100] was set pointing upwards, which is the positive $V$ axis in Fig.~\ref{fig:payload}. Different orientations may have different functionality advantages, but they are not evaluated in this work.

\begin{table}
\caption{ \label{tab:cases} Dimensions of the flexures analyzed. The shape of the flexure is shown in Fig.~\ref{fig:flexure-and-window}(a). The term {\em element} either refers to a flexure or to a suspension rod with a circular cross section and working in tension. The case name R-2 is used here for consistency with other documents \cite{flexure-design-fabian}.}
\begin{indented}
\lineup
\item[]\begin{tabular}{@{}lccc@{}}
\br 
\multicolumn{1}{l}{\begin{tabular}[l]{@{}l@{}}Case\\ name\end{tabular}}    & \multicolumn{1}{c}{\begin{tabular}[c]{@{}c@{}}Safety factor\\ (nominal)\end{tabular}} &  \multicolumn{1}{c}{\begin{tabular}[c]{@{}c@{}} Element \\ diameter (mm) \end{tabular}}  & \multicolumn{1}{c}{\begin{tabular}[c]{@{}c@{}} Element \\ length (mm) \end{tabular}} \\
\mr 
1-1 &   \03.1    &  \0\00.83  &   \0\01.7     \\
1-2 &   \03.1    &  \0\00.83  &   \0\03.4            \\
1-3 &   \03.1    &  \0\00.83  &   \05\phantom{.}            \\
2-1 &   \06.1    &  \0\01.16 &    \06\phantom{.}            \\
2-2 &   \06.1    &  \0\01.16 &    \09\phantom{.}            \\
2-3 &   \06.1    &  \0\01.16 &    12\phantom{.}            \\
3-1 &  18.1   &    2\phantom{.}   &    10\phantom{.}            \\
3-2 &  18.1   &    2\phantom{.}   &    24\phantom{.}            \\
3-3 &  18.1   &    2\phantom{.}   &    38\phantom{.}            \\
R-2  &  \06.1    &  \04.8   &    2000\phantom{.}\0\0   \\
\br 
\end{tabular}
\end{indented}   
\end{table}

A span of flexure cross sections with three different mechanical safety factors with respect to the strength of the material was considered. For each safety factor, three different lengths of the section between the nailheads were analyzed. The safety factors were calculated using the stress at the central cross section of the flexure \cite{flexure-design-fabian}, even though the maximum stress found by the numerical calculations occurs at the fillet and the material strength changes with different surface polish finishes and silicon orientations and grades. Therefore these values should be considered nominal.

Table \ref{tab:cases} shows the dimensions considered. The safety factors were 3.1, 6.1 and 18.1. The maximum lengths were chosen to be safe according to a buckling analysis made in Ansys \cite{flexure-design-fabian}. The longest flexures of each set were chosen to be  safely short of buckling.

\begin{figure}
	\centering
	\includegraphics{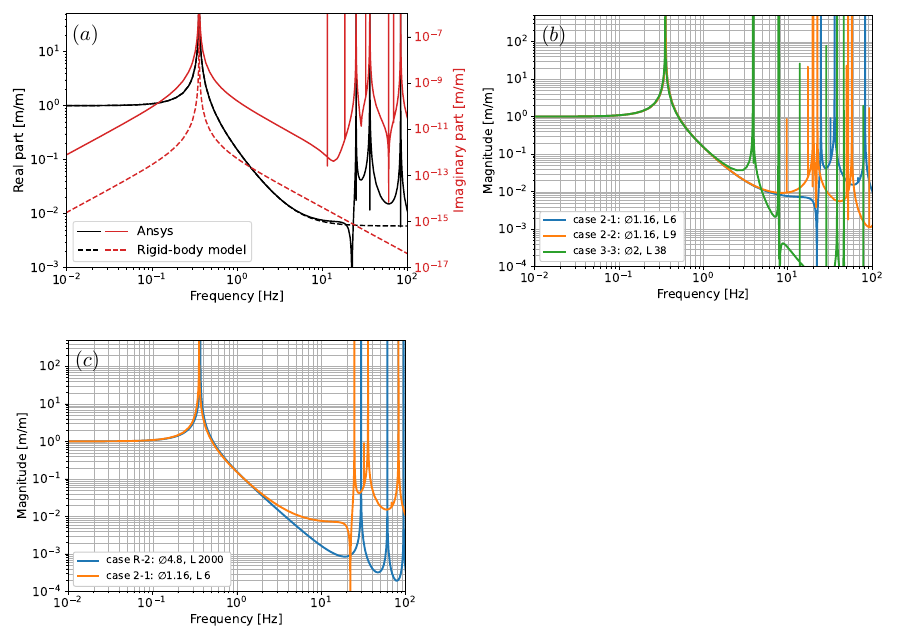}
	\caption{ \label{fig:tf}  Panel ($a$) shows the real and imaginary parts of the transfer function for the case 2-1, calculated with Ansys and with the rigid-body model that properly includes the effect of dissipation dilution. Panel ($b$) shows the transfer function calculated for three different flexure choices. The case of the flexure 3-3, which is safe from but close to buckling, presents a low frequency mode in which the beam moves rigidly, and most deformation happens in the upper flexure. Shorter flexures are stiffer, the equivalent of the rigid-body mode moves above the banana-shape mode of the beam, whose resonant frequency is only marginally affected by the flexure stiffness. Panel ($c$) compares the performances of cases 2-1 and R-2.}
 \end{figure}

In order to compare a couple of performance parameters, the case of a suspension rod with a circular cross section and working in tension was included in the list  with the name R-2. It has a safety factor of 6.1 assuming a  tensile strength of 165\,MPa \cite{azo-materials-si}. Using rods in tension is the traditional way of making suspensions that the configuration proposed in this paper aims to replace. Please note that for the calculations presented the rod ends have not been engineered as usual \cite{Cumming-2009}, and that the flexures are {\em safer} also because  in compression they are less sensitive to surface defects. 

In relevant cases, an analytic rigid-body model of the compound pendulum shown in Fig.~\ref{fig:flexure-and-window}(a) was used to compare with the flexible-body model from Ansys and COMSOL.  This model is based on the rigid-body model of an inverted pendulum \cite{Akiteru-PhD-2002,Blom-PhD-2015}. It takes into account both flexures and all the relevant mechanical properties of the beam, the counterweight and the load. The bending points of the flexures were assumed to be at the midpoints of their central sections. The bottom flexure was given a boundary condition of a horizontal top surface to mimic the situation of the mirror supported by four pendulum beams. The acceleration due of gravity was included, and the vertical displacements were expressed to second order in the angular displacement, leading the model to produce results already subjected to dissipation dilution \cite{CAGNOLI200039}, dispensing with the need of inserting the dissipation dilution factor later  in the results. The dimensions, mass and moment of inertia of the beam were read from the 3D-CAD used in the flexible-body simulation, and the value of the flexure spring constant used was calculated with Ansys and COMSOL as described in Sec.~\ref{sec:flexure-characterization}.

Section \ref{sec:mechanical-behaviour} describes the mechanical behaviour of the pendulum shown Fig.~\ref{fig:flexure-and-window}(a) for some representative flexure cases from Table \ref{tab:cases}. The properties of a flexure will be presented in Sec.~\ref{sec:flexure-characterization}. The behaviour of the pendulum also depends on the elastic properties of the suspension beam, whose influence is presented in Sec.~\ref{sec:beam}.

\subsection{Mechanical behaviour} \label{sec:mechanical-behaviour}

We consider the transfer function, in the longitudinal direction, from the intermediate mass to the mirror, for the pendulum in isolation as shown in Fig.~\ref{fig:flexure-and-window}(a),  with flexures from Table \ref{tab:cases}. The transfer function for the two-stage suspension was not evaluated as some mechanical relevant features of the upper stages are yet to be designed.

The transfer functions were calculated using Ansys. The anisotropic stiffness matrix of single crystal silicon \cite{Hopcroft-silicon} with direction [100] pointing upwards and a bulk loss angle $\phi_{\rm mat}=10^{-9}$ were used. To mimic the load of 1/4 of the mirror mass, a point mass of 50\,kg was positioned at the top face of the lower flexure and a counterweight of 0.65\,kg was placed at the upper end of the beam.

To get a sense for the mechanical loss, Ansys was also asked to compute the imaginary part of the transfer function. This is an upper bound on the true value, because programs like Ansys and COMSOL are known not to account correctly for dissipation dilution produced by gravity, as illustrated in Fig.~\ref{fig:tf}(a) and described below. In the case of the pendulum mode the dilution effect is large because the stored energy is mostly gravitational. In the case of violin modes in thin fibres, dilution happens because the sideways rigidity is created by the lengthwise tension, making the dissipated energy depend on the fourth power of the mode amplitude \cite{Mark-DD}. In the case of higher frequency modes of the thick suspension beam, dilution is not expected because the sideways rigidity is mostly produced by the intrinsic rigidity of the material, making the dissipated energy depend on the second power of the mode amplitude,  i.e., proportional to the stored energy in the mode, characteristic of systems without dilution.

For example, for the case 2-1, Fig.~\ref{fig:tf}($a$) shows the transfer function calculated in Ansys and with the rigid-body model in continuous and dashed lines respectively. The real and imaginary parts are shown, in black and red lines respectively, and each one with its own scale. The real parts coincide up to 20\,Hz, where the internal resonances of the suspension beam begin to appear. The imaginary parts, however, differ.  In Sec.~\ref{sec:thermal-noise}, using the rigid-body model, we show that the dissipation dilution factor is 313, so the true imaginary part of the transfer function in Fig.~\ref{fig:tf}($a$) is lower by this factor near the pendulum mode, and its quality factor will be higher by the same factor.

\begin{figure}
	\centering
	\includegraphics{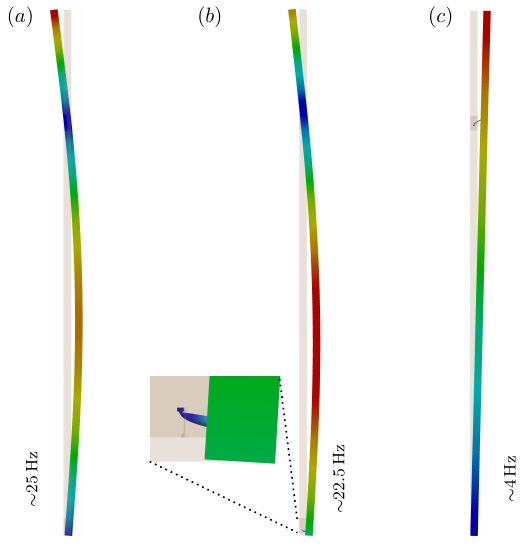}
	\caption{ \label{fig:two-modes}  	Panel ($a$) shows the banana mode for the case 2-1. Panel ($b$) shows a hybrid mode for the case 2-2, where the lower flexure experiences a larger deformation. Panel ($c$) shows a rigid-body mode for the case 3-3. The color scale goes from blue to red to indicate from the smallest to the largest deformation respectively. The counterweight mass is 0.65\,kg.}
 \end{figure}

Transfer functions for selected cases from Table \ref{tab:cases} are shown in Fig.~\ref{fig:tf}($b$), while the whole set is reported in Ref.~\cite{flexure-design-fabian}. A common feature is the pendulum mode at 356\,mHz, which is only marginally affected by the flexure stiffness. Each of the cases shown, however, exhibits distinctive behaviour at higher frequencies \cite{nelson-modes}. In the case 2-1, a banana-shape mode, illustrated in Fig.~\ref{fig:two-modes}(a), appears at around 25\,Hz.
In the case 2-2, whose flexure is longer, this mode is more affected by the flexure stiffness and the lower flexure experiences a larger deformation as shown in the detail view in Fig.~\ref{fig:two-modes}(b), where it is referred to as hybrid mode.
When the flexure is too slender, at frequencies below those of these banana and hybrid modes, other modes appear in which the beam behaves almost as a rigid body and most of the deformation happens at the upper flexure, as illustrated in Fig.~\ref{fig:two-modes}(c).

Case 2-1 was considered to be the most interesting for further calculations in this paper. Its flexure is 1.16\,mm in diameter and 6\,mm in length. The nailhead diameter is 3.48\,mm and the overall length is 11.8\,mm. Its mass is only 0.105 grams. It has the advantage of being short and thin, but still robust with a nominal safety factor of 6, and does not produce rigid-body modes below the banana mode besides the pendulum mode. This choice is to a certain extent arbitrary because the criteria for a better optimization are not available yet. 

Figure \ref{fig:tf}(c) compares the transfer functions of cases 2-1 and R-2, which have the same safety factors of 6. Above 20\,Hz the floor of the transfer function for the cylindrical rod is lower. As discussed in Sec.~\ref{sec:beam}, the plateau that appears below 20\,Hz is produced by the centre of percussion effect of the rigid beams and it can be lowered using the counterweight.

\subsection{Flexure mechanical characterization} \label{sec:flexure-characterization}

This section studies for the case 2-1, the values of the flexure angular spring constant, its loss angle and the maximum stress when bent to the maximum allowable deflection. In order to achieve an appropriate accuracy, these parameters were calculated with an idealized simplification of the single pendulum shown in Fig.~\ref{fig:flexure-and-window}(a). The beam was considered infinitely rigid and massless, the counterweight and the lower flexure were removed, and a point-mass load was attached to the beam 2\,m below the bottom face of the upper flexure. The point-mass load included the beam and the counterweight masses to produce the correct amount of compression on the flexure.

\begin{figure}
	\centering
	\includegraphics{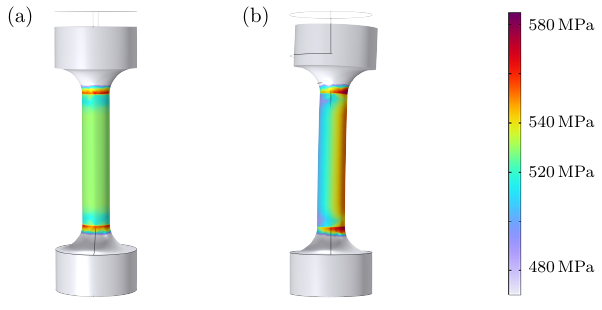}
	\caption{ \label{fig:flexure} Panel (a) shows the stress map of the flexure with a static vertical load. Panel (b) shows the map when the load has moved sideways to an end stop at ±5 mm from the equilibrium position, which represents a bend of 2.5\,mrad at the flexure. The color scale, adjusted to best illustrate the stress in the central section, is common to both cases. The deformation is exaggerated and wireframes correspond to the flexures without compression. The case 2-1 corresponds to a flexure with with \diameter 1.16\,mm and L\,6\,mm.}
 \end{figure}

The static stress without deflection across the 1.16\,mm diameter was calculated manually to provide a sanity check,
yielding a value of 525.2\,MPa. This was in good agreement with the value of 524.1\,MPa calculated in Ansys.
This stress was used to calculate the nominal safety factor, ignoring the stress peak found at the fillet joining the cylindrical section with the nailhead, which Ansys evaluated to be 556.3\,MPa.  The stress map is shown in Fig.~\ref{fig:flexure}(a). The maximum stress when the load is pushed to end-stops at $\pm$5\,mm from the equilibrium position was calculated to be  586.0\,MPa. The corresponding stress map is shown in Fig.~\ref{fig:flexure}(b). The stress in this condition remains compressive and, therefore, an  {\em effective safety factor} of 5.4 is maintained.

The flexure angular spring constant $k_{\theta}$ was calculated using Hooke’s law, without considering the effect of gravity to avoid including the gravitational spring constant.
In the Ansys simulation, transverse forces were  applied on the load, the lateral displacements were calculated and the value of the linear spring constant $k$ was derived from a least-square linear fit. The angular spring constant was then calculated from the expression $k_{\theta}=k l^2$, where $l=2\,$m. The length $l$ is far larger than the flexure length, which can be neglected. The values of the pendulum linear spring constant and of the flexure angular spring constant for case 2-1 are  $k=0.4\,$N/m and  $k_{\theta}=1.6\,{\rm Nm/rad}$ respectively.

\begin{figure}
	\centering
	\includegraphics{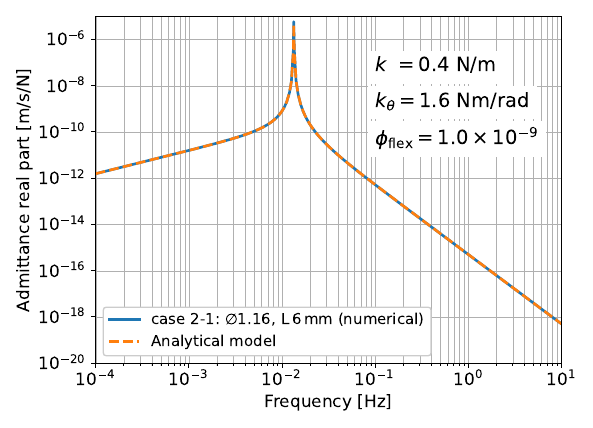}
	\caption{ \label{fig:flexure-characterization} The plot shows the mathematical model Eq.~\ref{eq:model-phi} fitted to numerically generated data with the goal of estimating the value of $\phi_{\rm flex}$.  The small value of the resonance frequency is produced by the absence of gravity in the calculation. The case 2-1 corresponds to a flexure with with \diameter 1.16\,mm and L\,6\,mm.}
 \end{figure}
 
 The loss angle of the flexure $\phi_{\rm flex}$ was estimated from the admittance of the pendulum calculated in Ansys. The admittance is a transfer function whose input is a force $F$ applied on the load and whose output is the velocity $V$ of the load. The analytical model of a harmonic oscillator with structural damping was fitted to the calculated admittance, procedure from which the value of  $\phi_{\rm flex}$ was estimated.
 The model is
\begin{equation}
\mathrm{Re}  \left( \frac{V}{F} \right) = \frac{ \omega \omega_0^2 \phi_{\rm flex} }{M \left[  \omega_0^4 \phi_{\rm flex}^2  + \left(   \omega_0^2 - \omega^2  \right)^2 \right]},
\label{eq:model-phi}
\end{equation}
where $M=56.6$\,kg including the beam and counterweight masses, $\omega_0=\sqrt{k/M}$ and $\omega$ is the angular frequency \cite[eq.~1.33]{Joris-PhD-thesis}. Fig.~\ref{fig:flexure-characterization} shows the numerical data, the fitted model and the resulting flexure loss angle $\phi_{\rm flex} = 10^{-9}$. This parameter characterizes the individual flexure only, and does not characterize the pendulum, whose own loss angle is subject to dilution with respect to $\phi_{\rm mat}$ due to the effect of gravity, which is absent from this calculation. In this case, $\phi_{\rm flex}=\phi_{\rm mat}$ because structural damping is the only dissipation mechanism considered.

There is a qualitative difference between a suspension rod in tension and a flexure in compression. In the case of the rod, the two effective bending points move progressively closer to the rod ends as the load  increases, making the bending length shorter.
 This effect is absent in flexures in compression that are thick enough to avoid buckling with a suitably large margin. Under these conditions, the flexures contribute uniformly to the deformation with their full length and are unaffected by load changes \cite{flexure-design-fabian}.

\section{ \label{sec:beam} Suspension beam design options}

The 2\,m length of the pendulum beam was selected to give a low pendulum frequency of 356\,mHz and reduce the suspension thermal noise. The length must also be optimized considering the compromises it entails, like longer beams providing better vibration attenuation but having internal resonance modes at lower frequencies, where astrophysical signals are expected. The size of the cryostat and the assembly procedures must be considered as well. These compromises were not analyzed in this conceptual study, but some will be referred to in the conclusions.

The beam thickness, in the direction that the displacement transfer function is calculated, was chosen considering the frequency of the lowest internal resonance that interrupts the  $1/f^2$ drop of the function, as illustrated in Fig.~\ref{fig:tf}($b$). The frequency of this mode increases linearly with the beam thickness \cite{nelson-modes}. 
A thickness of 35\,mm with frequency of 24\,Hz was selected for all the calculations presented here.  The thickness in the transverse direction was set to 25\,mm to limit the depth of the shelves machined at the sides of the mirror.

The maximum tensional stress in the beam occurs in the walls of the windows housing the flexures, with peak values of the order of about 1.5\,MPa \cite{flexure-design-fabian}, just a few percent of the maximum allowable tensional stress, offering a very large safety factor.

With these dimensions, the suspension beam mass is almost 5\,kg, far more than the fibres or rods considered for suspensions working in tension. Massive beams produce two harmful effects: the saturation of the displacement transfer function depicted in Fig.~\ref{fig:tf}, commonly referred to as the {\em centre of percussion effect} \cite{Akiteru-PhD-2002,Blom-PhD-2015}, and the appearance of beam resonances.

\subsection{Mitigation of the centre of percussion effect}

In order to mitigate the centre of percussion effect, the beam is equipped with an extension above the upper flexure whose length is 1/4 of the separation between the upper and the lower flexures, as depicted in Fig.~\ref{fig:flexure-and-window}(a). This length was selected also to make the extension work as a lever arm to actuate on the mirror by applying forces at its upper end.
 In an infinitely rigid suspension  beam, given that the position of the centre of percussion is fixed at the upper flexure, an appropriate counterweight mass can be chosen to position the centre of rotation at the lower flexure, thus avoiding the saturation of the displacement transfer function and preserving  a $1/f^2$ dependency to arbitrarily high frequencies.
For a flexible suspension beam, this correction extends only up to the perturbation of the first banana mode. There are different optimization strategies, whose usefulness depends on the specific requirements of the detector, which are not fully known yet.

\begin{figure}
	\centering
	\includegraphics{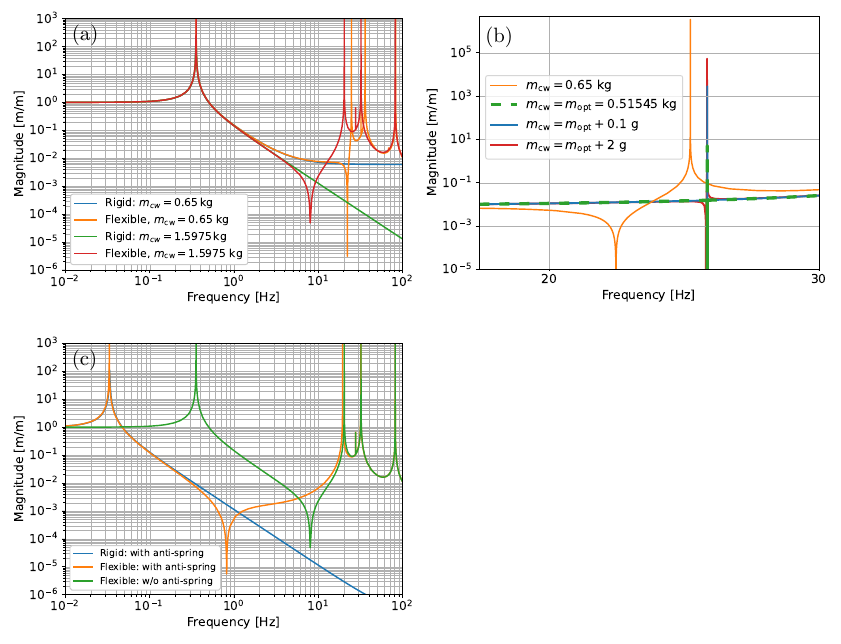}
	\caption{ \label{fig:tf-counterweight}  Displacement transfer functions for the case 2-1. Panel (a) shows a comparison between the rigid and flexible body models, and panel (b) shows the different amplitudes of the residual banana mode peak for different counterweight masses, where $m_{\rm opt} = 0.51545$\,kg. Panel (c) shows the effect of using an anti-spring to reduce the pendulum resonance frequency to  33\,mHz. For the three cases depicted  $m_{\rm cw}= 1.5975$\,kg. The case 2-1 corresponds to a flexure with with \diameter 1.16\,mm and L\,6\,mm.}
 \end{figure}

Fig.~\ref{fig:tf-counterweight}(a) shows some possible optimizations for the case 2-1.
The curve in green is the rigid-body transfer function. With an optimum counterweight mass of $m_{\rm cw}= 1.5975$\,kg, it exhibits an uninterrupted  $1/f^2$ dependency above the resonant frequency. This mass value was determined by trial an error upon visual inspection of the transfer function. The flexible-body transfer function should follow it as close as possible in order to maximize the vibration isolation. The red curve is the flexible-body transfer function with the same counterweight. They are similar up to 6\,Hz where a notch begins to grow.
The curve in blue is a saturated rigid-body transfer function calculated with $m_{\rm cw}= 0.56$\,kg. From 2\,Hz onwards it deviates from the $1/f^2$ dependency and at around 16\,Hz it settles at the $6\times10^{-3}$ m/m saturation level. The curve in orange is the transfer function calculated with the flexible-body model with the same counterweight. It coincides with the corresponding rigid-body function  up to 20\,Hz, followed by a notch just before the next resonant peak.

Another strategy is to tune the counterweight mass to  remove or reduce the amplitude of a resonant peak from the transfer function. For the case 2-1 also, Fig.~\ref{fig:tf-counterweight}(b) shows the transfer function with different amplitudes and widths of the peak above 24\,Hz for different values of this mass.  In orange  when $m_{\rm cw} = 0.65$\,kg, and in green when  $m_{\rm opt} = 0.51545$\,kg, with a significant reduction of the amplitude of the peak. 
Although complete removal of the peak is in principle possible, in practice it may be challenging to precisely adjust the counterweight masses in the assembly with the four suspension beams.  
The curves in blue and red show the effect on the peak when the mass is intentionally adjusted with a 100\,mg and a 2\,g error respectively. 

 Figure \ref{fig:tf-counterweight}(c) shows the transfer functions for the rigid and flexible-body models with a resonant frequency tuned to 33\,mHz with an anti-spring acting at the very top of the beam. The case without the anti-spring is also shown, and in all cases $m_{\rm cw}= 1.5975$\,kg. In Sec.~\ref{sec:optical-antispring} the implementation of an optical anti-spring will be proposed, but in all the calculations in this paper the anti-spring was simulated with a simple force following Hooke's law. The vibration attenuation between 50\,mHz and 6\,Hz is larger with the anti-spring.
The counterweight mass can be adjusted experimentally by directly measuring the saturation of the transfer function. This can be done with the sub-assembly depicted in Fig.~\ref{fig:flexure-and-window}(b). Performing this task without a load would amplify the displacement of the bottom of the pendulum and make it easier to measure, without affecting the saturation level at the frequencies of interest. 

\subsection{Mitigation of the effect of the suspension beam resonances on the mirror} \label{sec:beam-resonances}

The internal resonances of the suspension beams generate peaks in the detector sensitivity curve that are analogous to the violin modes of suspensions operating under tension.  
These resonances are typically excited by the impulsive forces needed to acquire lock of the mirror for the main interferometer.
Therefore, resonances in or near the observation band must be damped down to the environment cryogenic temperature or, preferably, be actively cooled below the gravitational-wave detector sensitivity.
 Considering the bulk loss angle  of silicon of $10^{-9}$,  peaks with quality factors of around $10^9$ are expected, value that implies very high amplitudes and lifetimes of months. In severe cases they can interfere with the locking of the interferometer via the feedback signal. 
During an observation run, the resonances are not expected to be easily excited, which means that any damping or cooling mechanism implemented can be safely switched off after use to avoid injecting noise. 

\begin{figure}
	\centering
	\includegraphics{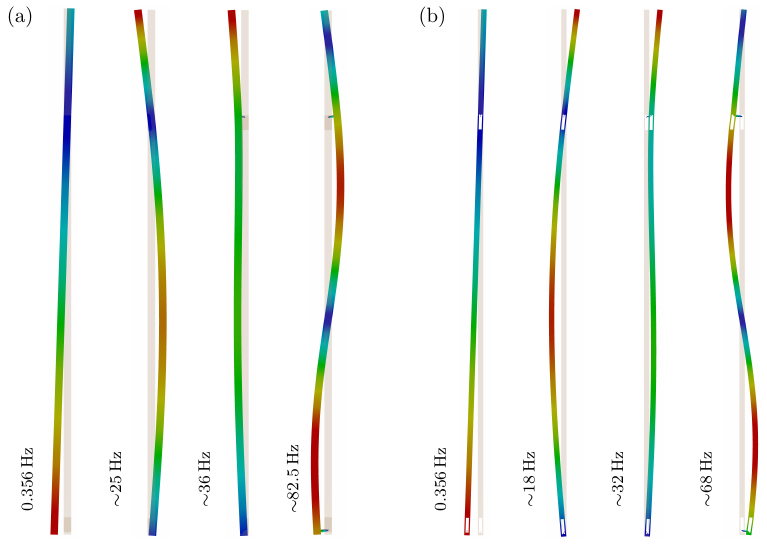}
	\caption{ \label{fig:modes-inventory}  For the case 2-1, the modes shown in panel (a) are longitudinal (along the optical beam line), and the ones shown in panel (b) are transverse. From left to right in each group, pendulum mode, banana mode, hybrid mode and S-shape mode. The color scale goes from blue to red to indicate from the smallest to the largest deformation respectively. The resonances at 82.5 and 68\,Hz are outside the low-frequency detector sensitivity. Other modes whose shapes are not depicted are an axial rotational mode at ${\sim}11.5$\,Hz and a vertical bounce mode at ${\sim}59.5$\,Hz. The case 2-1 corresponds to a flexure with with \diameter 1.16\,mm and L\,6\,mm.}
 \end{figure}

The beams offer a convenient geometry for damping the resonances either actively or passively. The damping or cooling are most effective when applied at the beam extension ends because, as illustrated in Fig.~\ref{fig:modes-inventory},  the resonances inside the detection band below 40\,Hz tend to produce amplitudes at least 30 times larger there than at the mirror \cite{nelson-modes}.  The beams also offer large flat side surface areas that can be used to implement displacement sensors and actuators. These areas are in close proximity of the intermediate mass where complementary parts of these devices can be mounted.

 Only single beam resonances are considered here.  In a more comprehensive analysis, with the mirror hanging from the four suspension beams, there will be degenerate modes, with different combinations of beams moving in phase or in phase opposition. While the resonance peaks may appear in the interferometer signal at very close frequencies to the point of being indistinguishable from one another, the motion of each beam is effectively independent from the others and can be passively or actively damped individually from the intermediate mass with the mechanism described in Sec.~\ref{sec:active-damping} below. 
 
Only longitudinal modes that directly affect the main interferometer readout are considered here. The transverse  modes mostly result in mirror rotation that affect the lateral stability of the main interferometer and are required to be damped too, but are less critical. 

\subsubsection{Passive damping of resonances.}

Resonant motion can be damped passively using  eddy currents. The required magnetic field can be generated with superconducting coils mounted on the intermediate mass, inducing  eddy currents in resistive pads deposited on the beam surfaces. 

A problem of using magnetic fields is that they couple with man-made magnetic fields, or even Earth’s magnetic field, which is shaken with $1/f$ noise by the solar wind even deep inside Earth’s crust. Once damping to the amplitude corresponding to the local temperature is achieved, the magnetic field can be turned off and with it the coupling to this noise. This is the reason for not advising the use of permanent magnets.

This type of damping is meant to be used briefly only after the lock acquisition of the main interferometer, which tends to excite the resonances, and should be kept  off during astronomical observations. Because of their large quality factor, the peaks may still stick above the detector noise curve even when damped to the local thermal bath temperature.

\subsubsection{Active cooling of resonances.} \label{sec:active-damping}

Passive damping is relatively easy to implement and may be adequate for certain modes, however, active cooling is preferable for longitudinal modes that directly affect the main interferometer readout because temperatures below the local thermal bath can be achieved reducing the peaks towards the detector sensitivity level. The active system would comprise interferometric displacement sensors and electrostatic actuators mounted on the intermediate mass and acting on the beam ends.  
A suitable optical sensor could be a fibre-fed, centimetre-long, high finesse Fabry-Perot interferometer.

The extent to which the peak amplitudes can be reduced will depend mostly on the sensitivity of the displacement sensor, and to a lesser extent on the design of the control filter. The resonant motion amplitude at the top end of the beams is at least 30 times larger than the amplitude at the test mass \cite{nelson-modes}. Therefore, the sensitivity of the displacement sensor would need to be 30 times the noise floor of the main interferometer to make the resonance peaks disappear.

Active damping can be used to cooldown the beam resonant modes after lock acquisition. When necessary, it can also be used during astronomical observations, but with a low gain to limit the local sensor noise that is injected into the mirror motion.

\subsection{Actuation for interferometer control without a recoil mass}

A recoil mass is a body that either surrounds the mirror or is in close proximity to it, from which actuation forces are applied. In the case of Virgo, the cage surrounding the mirror and the marionette serves as a recoil mass for both bodies \cite{Naticchioni-2018}, and is an extension of the vibration isolation chain rather than part of its payload. In KAGRA, the recoil mass of the mirror hangs from the recoil mass of the body above \cite{Ushiba-2021}.

In the case of the suspension presented in this paper, the suspension beams are rigid up to around 20\,Hz (see Fig.~\ref{fig:modes-inventory}); therefore, the same actuators used for cooling the internal resonances of the beams can be used to control the mirror to acquire and hold the lock of the main interferometer, thus dispensing with the need of a recoil mass around the mirror.
The forces required to act directly onto the mirror to acquire the lock can be calculated using standard formulas \cite[eq.~(6.6), p.~84]{Bersanetti-PhD-thesis}. For example, for a residual velocity between 0.1 and 1\,$\mu$m/s, a finesse of 400 and a wavelength of 1550\,nm, these forces are between 1 and 103\,mN. Therefore, at the position of the actuators at the top of the beams, due to the lever arm disadvantage of a factor of four,  the required forces range from 4 to 412\,mN. Because these forces will be generated by four actuators, the requirement for each is between 1 and 103\,mN. 

The longitudinal position of the mirror can be controlled by applying the same force on the four suspension beams, and yaw by applying forces with opposite signs on the left and right pair of beams. As will be also mentioned in Sec.~\ref{sec:spring-blades}, actuation in pitch is accomplished by applying forces of opposite sign to  the front and back pairs of blade springs.
Actuation on the angular degrees of freedom can be complemented by applying forces onto the intermediate mass using a recoil mass around it, which can either  hang from or be fixed to the frame of the upper mass. This recoil mass is not depicted in Fig.~\ref{fig:payload}. Another strategy is to apply forces to the bottom end of the central suspension beam, where a counterweight is shown in Fig.~\ref{fig:payload}.
Depending on the residual noise on the mirrors, it may even be that, after acquiring lock and damping the resonances, the lock can be maintained by radiation pressure alone using the Fabry-Perot displacement sensors implemented for modal cooling. In this case the feedback signal would come directly from the main interferometer.

Actuation on the mirror by applying a torque to the suspension beams will produce a reaction torque on the intermediate mass, which in turn can tilt the mirror. This effect could in principle be eliminated by diagonalizing the actuation along the length and pitch degrees of freedom for the mirror and intermediate mass \cite[p.~151]{Beker-PhD-thesis}.

Interferometer control without a recoil mass is an invaluable simplification and provides a drastic reduction of the mass of the cryogenic payload. Still, around the mirror there must other components, like a security structure and baffles for scattered light reduction, which can be fixed or hung from the cryostat or structures other than the suspension.

\subsection{Pendulum frequency adjustment with an optical anti-spring} \label{sec:optical-antispring}

The resonant frequency of a passive pendulum is determined by its length, with longer lengths producing lower frequencies. This frequency can be reduced actively with a tuneable anti-spring produced by a fibre-fed detuned Fabry-Perot cavity working over a limited displacement range, acting in parallel with the gravitational restoring force. As illustrated in Fig.~\ref{fig:tf-counterweight}(c), the relevance of this strategy is that it will improve the mechanical attenuation.
As in the case of the local displacement sensors and actuators, the optical anti-spring would be placed at the very top of each of the four suspension beams holding the mirror, with their complex stabilization and control optics remaining outside of the vacuum chamber. 

In a Fabry-Perot cavity, when its end mirror reflectance equals unity, the optical spring constant of the radiation pressure force acting on the input mirror is \cite{DiPace2020}
\begin{equation}
k_{\rm opt} = - \frac{64 \omega_0 I_0}{c^2 T_1^2} \frac{\delta_\gamma}{ \left( 1+\delta_\gamma^2 \right)^2 },
\end{equation}
where $\omega_0$, $I_0$ and $c$ are the angular frequency, input power and speed of light respectively,  $T_1$ is the input mirror transmittance, $\delta_\gamma$ is the normalized detuning parameter 
\begin{equation}
\delta_\gamma = \frac{4 L}{c T_1} \left(  \omega_{\rm r} - \omega_0  \right),
\end{equation}
and $L$ and $\omega_{\rm r}$ are the length and  the resonance angular frequency of the cavity.
If we keep $\omega_{\rm r}$ fixed by keeping $L$ fixed, the size and sign of the optical spring constant is determined by the variation of  the wavelength $\lambda$  via $\omega_0 = 2 \pi c / \lambda$. 

As an example, for $\lambda = 1064$\,nm, $I_0=3.9$\,mW, $L=1$\,cm, $T_1=3.21 \times 10^{-4}$ and finesse $\mathcal{F}= 9823$, these formulas yield $k_{\rm opt} = -3902$\,N/m for a wavelength variation $\Delta \lambda = 1.7 \times 10 ^{-15}$\,m ($\Delta f_0 = \Delta \omega_0 / 2 \pi = 414$\,kHz). The parameter $k_{\rm opt}$ varies within 10\% over 15.6\,pm \cite{selleri-report}. A setup like this can compensate for the stiffness of 3901\,N/m at the top of each of the suspension beams calculated with the rigid-body model, which includes the effect of the masses of the beam and of the counterweight.

A working displacement range of 15.6\,pm may seem sufficient considering that the residual motion achievable at the bottom of the seismic attenuation chain is far smaller in the frequency range of astronomical observations. However, at very low frequencies, where the attenuation chain is not effective, the residual motion is larger. Therefore, taking into account the performance of the chain, it is still necessary to estimate the relative displacement between the intermediate mass and the upper beam end,  and adjust the displacement working range accordingly.  For example, more optical power and a lower finesse would allow to increase it. Another option would be to implement this active anti-spring with an optical displacement sensor and an electrostatic actuator.

\section{Blade spring design} \label{sec:spring-blades}

The main role of the blade springs shown in Figs.~\ref{fig:payload} and \ref{fig:payload-details} is to provide the vertical compliance for compensating the length manufacturing tolerances of the four beams that hold the mirror. They also offer a means to actuate on the pitch of the mirror by applying forces of opposite sign to the  front and back pairs of blade springs.

They have an isosceles triangular shape, with the load positioned near the intersection point of the two equal sides of the triangle. Their length is constrained by the radius of the mirror and their width by the separation between the front and back suspension beams. The tips of the blades have straight sections on their upper faces for supporting the cross beams and for applying pre-loading forces before placing the final load.

As illustrated in Fig.~\ref{fig:payload-details}(b), the blades are inserted into slots in the central block of the intermediate mass and held by the torque produced by the load. The profile of the lower surface of the slot must have a curvature varying from zero to that of the loaded blade to produce a smooth transition of the stress along the supporting face of the blade. The slots have a tilt angle to make the blade tips horizontal when loaded. The pre-assembled sub-assemblies shown in Fig.~\ref{fig:flexure-and-window}(b)  can be attached to the rest of the suspension inside the cryostat with gallium brazing to avoid heating to temperatures much higher than room temperature, and for easy disassembly when needed. 
 
\begin{figure}
	\centering
	\includegraphics{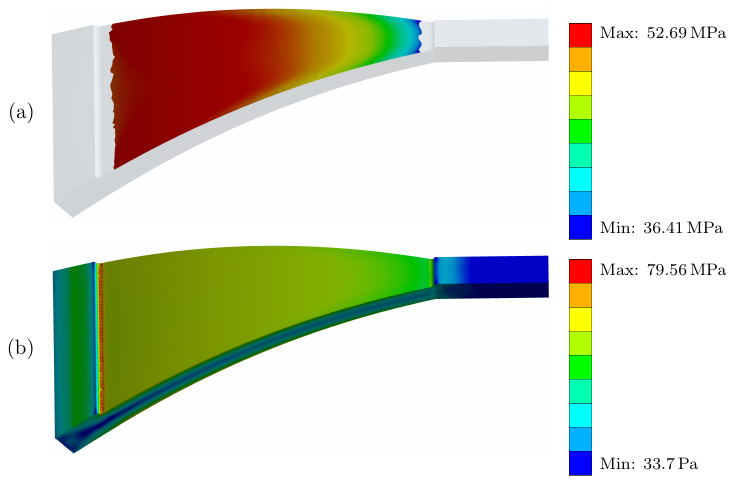}
	\caption{ \label{fig:blade} (a) Tensile stress map of the upper face of the loaded blade, showing that a relatively uniform stress distribution is achieved  with  the triangular shape, (b) stress map for the blade showing that the highest stress occurs at the fillets.}
 \end{figure}
 
Fig.~\ref{fig:blade} shows a 10.5\,mm thick loaded blade with a greatly exaggerated deformation. The actual value of the radius of curvature is 13.63\,m, the base tilt 12.5\,mrad  and the sag of the tip 1.3\,mm when loaded.
The base and the tip are thicker at the upper face to reduce the tensional stress in the attachment areas. The upper face experiences an almost uniform tensional stress with a maximum value of 52.69\,MPa. The overall maximum was estimated with Ansys to be 79.55\,MPa, occurring along the fillets at the transitions to the thicker sections. It can be reduced by designing fillets with an optimal shape.

The thickness was chosen to maintain a safety factor of 3.1 considering the tensional breaking strength of 165\,MPa \cite{azo-materials-si}. Thinner and more stressed blades can be designed if tests show that a higher stress is tolerable with a good quality surface finish.

The spring constant of this blade is  $k=330$\,kN/m \cite{leonardo-report}. Assuming an accumulated manufacturing error in the length of the beams and, therefore, in the position of the attachment faces of 0.1\,mm, the load on the four beams will differ by 33.0\,N which is 6.7\% of what each blade would support when the load is equally shared.

With a vertical resonant frequency of 13.6\,Hz, the blades cannot provide any useful passive vertical attenuation, which is needed to filter out the vertical thermal noise of the seismic attenuators above. Due to Earth’s radius, the test masses will be tilted by at least 0.78\,mrad with respect to their respective local plumb lines that are 10\,km apart. Therefore, at least 0.08\% of the vertical thermal noise will be projected into the detector length noise. This is a serious limitation for the sensitivity at low frequencies because the vertical filters of the vibration isolation chain have metallic springs that produce large vertical thermal noise. Therefore, to achieve better attenuation it is necessary to implement active anti-springs similar to the one proposed for the pendulum beams \cite{STOCHINO2009737}. The problem is that the stiffness of the blade springs is almost two orders of magnitude larger than that of the pendulum at the top of the beam, which is beyond the force authority of an optical anti-spring. Electrostatic actuators slaved to Fabry-Perot displacement sensors can be used to provide more force authority.  In a parallel plate actuator with surface $A = 4$\,cm$^2$, a separation $d=0.3$\,mm, operated up to a voltage $V=1000$\,V, the maximum force provided is $F=20$\,mN. This authority allows a maximum range of 60\,nm, which is sufficient at the bottom of a seismic attenuation chain where the residual mechanical noise is measured in picometers.  
Each of the four blade springs will be equipped with an electrostatic actuator, and they will be operated in parallel as a single anti-spring.

One of the relevant contributions to the noise budget of the softened blade springs is the noise from the Fabry-Perot displacement sensors introduced via the feedback  loops. Another one is their own thermal noise. Although lowering the vertical resonance frequency of the blade springs improves their vibration attenuation capabilities, it does not decrease their own thermal noise. As shown in Fig.~\ref{fig:thermal-noise}(c) for an anti-spring for the suspension beams, the thermal noise increases below the open-loop resonant frequency. This happens because the Brownian noise force remains the same while the mechanical response of the system increases. Therefore, there is a limit to the attenuation of noise coming from upper stages given by the thermal noise of the blade springs themselves and the noise introduced when using the anti-spring.

Despite this limit, the benefit of actively softening the blade springs could be large. The attenuators in the chain use blade springs made of maraging steel with a loss angle of  ${\sim}10^{-4}$ operating at 300\,K, whereas the blades are made of silicon with a loss angle of  ${\sim}10^{-9}$ operating below 10\,K. The attenuation of the noise coming from upper stages could be substantial at the optimal limit in which the residual unfiltered noise equals the softened blade spring thermal noise.

\section{ \label{sec:temperature} Mirror operation temperature and cooling time}

In this section, the thermal performance of the suspension is quantified with three parameters: the time it takes to cool the mirror down from room temperature to 4\,K, the time it takes for the cold mirror to reach a steady temperature after the main interferometer has been locked, and that temperature. These parameters were calculated imposing a 4\,K temperature to four spots in the octagonal aluminium frame of the upper mass shown in Fig.~\ref{fig:payload}. 

 Despite their short length, due to their small diameters, the flexures set the limit to the suspension thermal conductance. The performance can improve significantly by using flexures made with isotopically pure silicon $^{28}$Si, which exhibits a much larger thermal conductivity due to the absence of $^{29}$Si and $^{30}$Si atoms, which scatter the heat-carrying phonons and thus reduce their mean-free path. Below 100\,K, the thermal conductivity of  $^{28}$Si is higher than for natural silicon with a peak 10 times higher at 20\,K \cite{10.1063/1.5017778}. The specific heat capacity used in the calculations for both types of silicon was taken from Ref.~\cite{okhotin-si}. It is important to note that the radiative cooling, which is significant above 100\,K, was neglected in the calculations, and therefore the actual cooling time from room temperature can be expected to be shorter.

\begin{figure}
	\centering
	\includegraphics{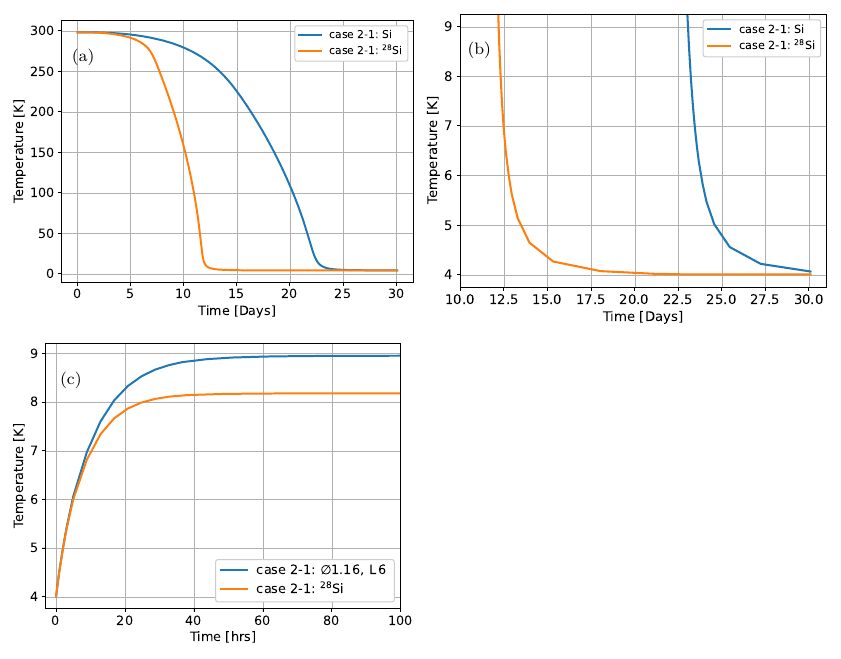}
	\caption{ \label{fig:si-thermal} Panel (a) shows the mirror temperature during cooldown with natural silicon and isotopically pure $^{28}$Si. Panel (b) shows part of the same with narrower temperature and time ranges. Panel (c) shows the mirror temperature when the interferometer has been locked at the beginning of the time interval and 0.5\,W of light is dissipated in it. The case 2-1 corresponds to a flexure with \diameter 1.16\,mm and L\,6\,mm.}
 \end{figure}

 For the case 2-1, the temperature evolution during cooldown is plotted in Fig.~\ref{fig:si-thermal}(a) and (b). For flexures made of Si and $^{28}$Si, it takes about 25 and 14 days respectively to reach a temperature of 5\,K. The time evolution at different points of the suspension is shown in a separate report \cite{flexure-design-fabian}.
 Fig.~\ref{fig:si-thermal}(c) shows the temperature evolution of the mirror when the system is already cold and  the main interferometer is locked.
When the flexures are made of natural silicon, the mirror settles at a temperature of 8.9\,K  in about 44 hours, and when the flexures are made of  $^{28}$Si, the mirror reaches 8.2\,K in 33 hours. These operational temperatures are substantially lower than most expected from suspension rods working in tension \cite{PhysRevD.108.123009}, a condition that is also beneficial for achieving lower coating thermal noise.  
 
\section{ \label{sec:thermal-noise} Suspension thermal noise}

As in the case of the transfer functions reported in Sec.~\ref{sec:mechanical-behaviour}, in this initial study the suspension thermal noise was calculated for the single pendulum shown in Fig.~\ref{fig:flexure-and-window}(a). Three additional simplifications were made: only the bulk contribution of the material to the loss angle was considered, with a value of $\phi_{\rm mat} = 10^{-9}$; that this value is constant in temperature; and that the suspension is in thermal equilibrium. A more realistic estimate of the loss angle must take into account surface losses and temperature dependence that cannot be evaluated without experimental data. Also, the presumably lower loss angle of $^{28}$Si was not considered because it is still unknown. The thermoelastic noise is expected to be negligible at the low frequencies of the astronomical observations given the high thermal conductivity and the small diameter of the flexures. The brazing losses are expected to be small too \cite{PhysRevLett.132.231401}. Also, after the main interferometer has been locked and the suspension has reached a steady state, the temperature gradient produced by the heat flow results in a nonequlibrium condition in which different flexures are at different temperatures, a condition that needs to be taken into account \cite{PhysRevD.97.102001}. All these elements may be considered once more data becomes available.

The displacement thermal noise in the frequency domain was calculated using the expression
\begin{equation}
S = \sqrt{ \frac{4 k_B T}{\omega^2} \mathrm{Re}  \left( \frac{V}{F} \right)  }, \label{eq:thermal-noise} \label{eq:suspension-thermal-noise}
\end{equation}
where $k_B$ is the Boltzmann constant, $T$ the temperature, $\omega$ the angular frequency,  and the ratio $V/F$ is the admittance transfer function as a function of $\omega$, where $F$ is a force applied on the load and $V$ is the velocity of the load \cite{PhysRevD.57.659}. The rigid-body model was used in the calculation for its  ability to correctly account for dissipation dilution due to gravity without having to separately introduce a correction factor in the result, as opposed to commercially available finite-element analysis software \cite{Mark-DD,flexure-design-fabian}.

Dissipation dilution is the phenomenon in which the loss angle of the pendulum $\phi_{\rm pen}$ is smaller than that of the material $\phi_{\rm mat}$ due to the kinetic energy being also stored in the conservative gravitational field, rather than entirely becoming elastic potential energy stored in the dissipative material \cite{CAGNOLI200039}. It is quantified as
\begin{equation}
\phi_{\rm pen} = \frac{ \phi_{\rm mat } } {D},
\end{equation}
where the dissipation dilution factor $D$ is defined as the ratio of the gravitational and elastic potential energies, and can also be expressed as $D = k_{\rm grav} / k_{\rm assy }$, where $k_{\rm grav}$ and $k_{\rm assy}$  are the spring constants produced by gravity and by the elastic pendulum assembly respectively. 
For the spring constants, the rigid-body model gives $k_{\rm grav}=256.44\, {\rm N/m}$ and $k_{\rm assy} = 0.819 \, {\rm N/m }$, which implies that $D=313.1$. Therefore, the expected loss angle of the pendulum is $\phi_{\rm pen} = 3.19 \times 10^{-12}$. The flexible-body model gives  256.61 and 0.824\,N/m for  $k_{\rm grav}$ and $k_{\rm assy}$ respectively.
   
\begin{figure}
	\centering
	\includegraphics{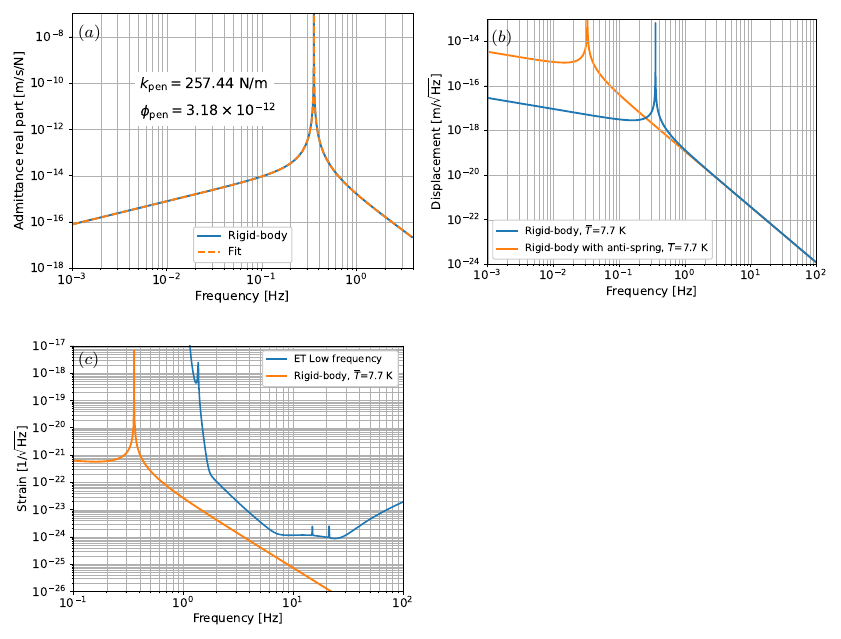}
	\caption{ \label{fig:thermal-noise} Panel ($a$) shows the admittance transfer function, which yields a value of the pendulum loss angle $\phi_{\rm pen}$ equal to the one calculated independently using  the dissipation dilution factor $D$.  Panel ($b$) shows the suspension thermal noise without and with the anti-spring. Panel ($c$) shows a comparison with the ET low-frequency sensitivity design.  All panels correspond to the case 2-1, with a flexure with \diameter 1.16\,mm and L\,6\,mm.}
 \end{figure}

The value of $\phi_{\rm pen}$ can also be extracted from the real part of the admittance, calculated with the  rigid-body model and shown in  Fig.~\ref{fig:thermal-noise}($a$) for the case 2-1.  As done in Sec.~\ref{sec:flexure-characterization} for the case of a single flexure, the mathematical model (\ref{eq:model-phi}) was fitted to the real part of the admittance, substituting the name of the the free parameter $\phi_{\rm flex}$ by $\phi_{\rm pen}$. The fitted value was $\phi_{\rm pen} = 3.18 \times 10^{-12}$, which is close enough to the expected one, suggesting that the calculation is correct. 

Figure \ref{fig:thermal-noise}($b$) shows the suspension thermal noise that follows from formula (\ref{eq:suspension-thermal-noise}). The temperature $\overline{T}$ used in the formula is the average temperature of the pendulum. Figure \ref{fig:thermal-noise}($c$) depicts the noise in units of strain considering the length of the arms of the interferometer and the contributions of the four independent mirrors added in quadrature.
The desired sensitivity of the low-frequency interferometer of the Einstein Telescope is shown for comparison. In this rigid-body approximation, the thermal noise of the suspension is lower than such a sensitivity. A more accurate comparison requires considering the two-level suspension and that the suspension beams are flexible and experience internal resonances.

Figure \ref{fig:thermal-noise}($b$) shows the effect of the optical anti-spring on the suspension thermal noise. As in Fig.~\ref{fig:tf-counterweight}(c), the pendulum resonant frequency was tuned to 33\,mHz. The $1/f^2$ slope extends to lower frequencies down to the new resonance frequency peak. This happens because the pendulum is softer at low frequencies, and, therefore, undergoes larger displacements under the effect of the thermal bath radiation that participates in the production of the thermal noise. 
Nevertheless, this effect is out of the frequency band of astrophysical observations and is, therefore, harmless.

\section{Conclusion}

This paper presents a conceptual analysis of a suspension with flexures working in compression for the cryogenic test-mass mirrors of the Einstein Telescope.  This research was guided by the necessity of a suspension with low thermal noise and high thermal conductance.
Its key elements are the flexures, which work in compression and are thin and short in order to be soft and also allow efficient heat extraction.
They work as hinges for long and rigid vertical suspension beams, which have large cross sections that results in low tensional stress and high thermal conductance.

Different flexure dimensions were analyzed and one case was selected for subsequent studies based on the frequencies of resonant modes and the value of the mechanical safety factor.  The choice was reasonable but a true optimization can only be made based on a holistic view of the entire gravitational-wave detector.

This innovative configuration is expected to be easy to implement  and is competitive with respect to all other solutions proposed so far \cite{PhysRevD.108.123009}. Its distinctive features, most of which are also novel, were only partially analyzed and are listed below together with future research and development opportunities. Not all these features need to be implemented from the beginning, but they can be added at different stages of development:

\begin{itemize}
	\item The implementation is flexible, thanks to the ease of assembly and disassembly using simple, separately machined components. The assembly can be be carried out inside of the cryostat. Components of different sizes could be used in different stages of the detector development.
	\item Actuation on the mirror using the rigid suspension beams as levers to acquire and maintain the lock of the main interferometer eliminates the need of a recoil mass.
	\item The frequency of the pendulum can be lowered with optical anti-springs without increasing its length to improve the vibration isolation of the test-mass mirror.
	\item Active softening of the vertical blade springs using fibre-fed optical length sensors and electrostatic actuators can mitigate the residual vertical noise, either thermal or seismic. 
	\item  Similar active systems can be used to cool the resonances below the temperature of the thermal bath and below the detector sensitivity level.
	\item The design allows taking advantage of the exceptional thermal and dissipation properties of isotopically pure silicon $^{28}$Si for the tiny flexures at an affordable cost. These flexures reduce the cooling time from room temperature to about two weeks and the test mass operation temperature to 8.2\,K as shown in Fig.~\ref{fig:si-thermal}. If the $^{28}$Si quality factor turns out to be larger than the $10^9$ value we assumed for natural silicon, the thermal noise, which is shown in Fig.~\ref{fig:thermal-noise}($c$) to be already below the ET desired low-frequency sensitivity for a single pendulum, will be further reduced.

\end{itemize}

The new concept will affect the design of other components of the facility. For example, longer pendulums will be needed in the seismic attenuation chain above to take full advantage of the lower frequencies achievable in the suspensions. The additional attenuation that the suspension provides may  allow a reduction in the number of seismic filters in the chain. The longer pendulum length and the longer spacing in between the components would allow the use of small access tunnels at multiple levels along the chain rather than a large hall housing the whole chain in a single space as in Virgo \cite{app12178827}. This strategy would allow  a drastic  reduction of the underground hall sizes, leading to a lower excavation cost.

The design was based on silicon because it is softer, but  sapphire mirrors can also be supported by the same silicon suspensions as long as the lower cross beams are made of sapphire with the same crystal orientation of the mirror to avoid the stress induced by the differential thermal expansion between the three contact areas of the mirror during cooldown, which would inevitably deform it beyond specifications. The contact area between silicon and sapphire would be limited to the small head diameter of the lower flexures with no effect on the mirror shape. 

As far as versatility is concerned, the stiff but actively softened vertical blade springs allow for suspending mirrors of different masses without physical changes as long as the mechanical safety margins of blades and flexures are respected.
Ballast masses on the upper mass can be used to readjust the load of the seismic attenuation chain above when the mirror mass changes.

An important point is that unless the large vertical thermal noise generated in the seismic attenuation chain is filtered out, it could be a limiting factor for the ET low-frequency sensitivity to gravitational waves. 
Currently, there is no known passive configuration with low thermal noise  that can produce the necessary mechanical attenuation, even when using  a material exhibiting low structural dissipation like silicon.
Because of this, the use of active anti-springs to reduce the vertical resonant frequency of the blade springs seems like a necessity, although they are one of the most demanding component proposed in this work. We consider that combining the optical length sensing and the electrostatic actuation outlined here could solve the problem, but this development will require a substantial effort.

\section{Acknowledgments}

We acknowledge the support from California State University Los Angeles NSF award 2309294.
At the University of Guadalajara, this work was in part supported by CONAHCyT Network Project No.~376127, entitled {\it Sombras, lentes y ondas gravitatorias generadas por objetos compactos astrofísicos}. CM wishes to thank PROSNI and Fondos Concurrentes, CUCEI, University of Guadalajara and CONAHCyT. FEPA wishes to express his gratitude to Prof.~Masatake Ohashi from the KAGRA Observatory for his support, and acknowledges receiving the scholarship {\it Beca de repatriación}  from CONAHCyT during the academic year 2023-2024.
We also acknowledge the financial support of the INFN (Italian National Institute for Nuclear Physics) in the framework of the ET-Italia project.
RDS acknowledges the input from Adalberto Giazotto who detailed the challenging cryogenic suspension requirements. It took 25 years, and the work of many, to find these feasible and effective solutions.



\section*{References}
\begin{thebibliography}{99}
\normalsize 

{}
\bibitem{Maggiore-2020}
Michele Maggiore et al. “Science case for the Einstein Telescope”. In:
\emph{Journal of Cosmology and Astroparticle Physics} 2020.03 (2020), p. 050.
\textsc{doi}: \href {https://doi.org/10.1088/1475-7516/2020/03/050} {\nolinkurl
{10.1088/1475-7516/2020/03/050}}. \textsc{url}: \url
{https://dx.doi.org/10.1088/1475-7516/2020/03/050}.
{}
\bibitem{ET-feasibility}
Eintein Telescope Steering Committee et al. \emph{Einstein Telescope: Science
Case, Design Study and Feasibility Report}. Technical Report ET-0028A-20.
Available from \url { https://apps.et-gw.eu/tds/ql/?c=15662 }. 2020.
{}
\bibitem{PhysRev.83.34}
Herbert B. Callen and Theodore A. Welton. “Irreversibility and Generalized
Noise”. In: \emph{Phys. Rev.} 83 (1 1951), pp. 34–40. \textsc{doi}: \href
{https://doi.org/10.1103/PhysRev.83.34} {\nolinkurl {10.1103/PhysRev.83.34}}.
\textsc{url}: \url {https://link.aps.org/doi/10.1103/PhysRev.83.34}.
{}
\bibitem{Kubo-1966}
R Kubo. “The fluctuation-dissipation theorem”. In: \emph{Reports on Progress
in Physics} 29.1 (1966), p. 255. \textsc{doi}: \href
{https://doi.org/10.1088/0034-4885/29/1/306} {\nolinkurl
{10.1088/0034-4885/29/1/306}}. \textsc{url}: \url
{https://dx.doi.org/10.1088/0034-4885/29/1/306}.
{}
\bibitem{PhysRevD.57.659}
Yu. Levin. “Internal thermal noise in the LIGO test masses: A direct approach”.
In: \emph{Phys. Rev. D} 57 (2 1998), pp. 659–663. \textsc{doi}: \href
{https://doi.org/10.1103/PhysRevD.57.659} {\nolinkurl
{10.1103/PhysRevD.57.659}}. \textsc{url}: \url
{https://link.aps.org/doi/10.1103/PhysRevD.57.659}.
{}
\bibitem{ET-2020-report}
ET Steering Committee. \emph{ET design report update 2020}. Official
Document ET-0007C-20. Available from \url {
https://apps.et-gw.eu/tds/ql/?c=15418 }. 2020.
{}
\bibitem{halliday2013fundamentals}
D. Halliday, R. Resnick, and J. Walker. \emph{Fundamentals of Physics
Extended, 10th Edition}. Wiley, 2013. \textsc{isbn}: 9781118473818.
\textsc{url}: \url {https://books.google.com.mx/books?id=DTccAAAAQBAJ}.
{}
\bibitem{azo-materials-si}
\emph{A Background to Silicon and its Applications}. \url
{https://www.azom.com/properties.aspx?ArticleID=599}. Accesed on 04-09-2024.
{}
\bibitem{Blom-PhD-2015}
Mathieu Blom. “Seismic attenuation for Advanced Virgo Vibration isolation for
the external injection bench”. PhD thesis. Vrije Universiteit Amsterdam, 2015.
\textsc{url}: \url
{https://www.nikhef.nl/pub/services/biblio/theses_pdf/thesis_M_Blom.pdf}.
{}
\bibitem{VIEIRA2024127549}
Lucas Vieira et al. “Simulation of crucible-free growth of monocrystalline silicon
fibres for mirror suspension in gravitational-wave detectors”. In: \emph{Journal
of Crystal Growth} 629 (2024), p. 127549. \textsc{issn}: 0022-0248. \textsc{doi}:
\href {https://doi.org/https://doi.org/10.1016/j.jcrysgro.2023.127549}
{\nolinkurl {https://doi.org/10.1016/j.jcrysgro.2023.127549}}. \textsc{url}: \url
{https://www.sciencedirect.com/science/article/pii/S002202482300475X}.
{}
\bibitem{ET-materials-database}
\emph{Einstein Telescope Materials Database}. \url
{https://wiki.et-gw.eu/ISB/MaterialsDatabase/WebHome?validation_key=f60217d8327cd76bc5812648910c6168}.
Accesed on 08-10-2024.
{}
\bibitem{Akiteru-PhD-2002}
Akiteru Takamori. “Low Frequency Seismic Isolation for Gravitational Wave
Detectors”. PhD thesis. University of Tokyo, 2002. \textsc{url}: \url
{https://granite.phys.s.u-tokyo.ac.jp/takamori/thesisver2002.pdf}.
{}
\bibitem{Nelson-master-thesis}
Nelson Leon. Work in progress. MA thesis. California State University at Los
Angeles.
{}
\bibitem{STOCHINO2009737}
Alberto Stochino et al. “The Seismic Attenuation System (SAS) for the Advanced
LIGO gravitational wave interferometric detectors”. In: \emph{Nuclear
Instruments and Methods in Physics Research Section A: Accelerators,
Spectrometers, Detectors and Associated Equipment} 598.3 (2009), pp. 737–753.
\textsc{issn}: 0168-9002. \textsc{doi}: \href
{https://doi.org/https://doi.org/10.1016/j.nima.2008.10.023} {\nolinkurl
{https://doi.org/10.1016/j.nima.2008.10.023}}. \textsc{url}: \url
{https://www.sciencedirect.com/science/article/pii/S0168900208015064}.
{}
\bibitem{10.1063/1.5017778}
Alexander V. Inyushkin et al. “–Ultrahigh thermal conductivity of isotopically
enriched silicon˝”. In: \emph{Journal of Applied Physics} 123.9 (Mar. 2018),
p. 095112. \textsc{issn}: 0021-8979. \textsc{doi}: \href
{https://doi.org/10.1063/1.5017778} {\nolinkurl {10.1063/1.5017778}}. eprint:
\url
{https://pubs.aip.org/aip/jap/article-pdf/doi/10.1063/1.5017778/19756584/095112“˙1“˙online.pdf}.
\textsc{url}: \url {https://doi.org/10.1063/1.5017778}.
{}
\bibitem{PhysRevLett.132.231401}
Karen Haughian et al. “Temperature Dependence of the Mechanical Dissipation
of Gallium Bonds for Use in Gravitational Wave Detectors”. In: \emph{Phys.
Rev. Lett.} 132 (23 2024), p. 231401. \textsc{doi}: \href
{https://doi.org/10.1103/PhysRevLett.132.231401} {\nolinkurl
{10.1103/PhysRevLett.132.231401}}. \textsc{url}: \url
{https://link.aps.org/doi/10.1103/PhysRevLett.132.231401}.
{}
\bibitem{flexure-design-fabian}
Fabi\'an Pe\~na Arellano et al. \emph{A first approach to flexure design for a
cryogenic test mass suspension}. Technical Report ET-0038A-25. Available from
\url { https://apps.et-gw.eu/tds/ql/?c=17765 }. 2025.
{}
\bibitem{Cumming-2009}
A Cumming et al. “Finite element modelling of the mechanical loss of silica
suspension fibres for advanced gravitational wave detectors”. In: \emph{Classical
and Quantum Gravity} 26.21 (2009), p. 215012. \textsc{doi}: \href
{https://doi.org/10.1088/0264-9381/26/21/215012} {\nolinkurl
{10.1088/0264-9381/26/21/215012}}. \textsc{url}: \url
{https://dx.doi.org/10.1088/0264-9381/26/21/215012}.
{}
\bibitem{CAGNOLI200039}
G. Cagnoli et al. “Damping dilution factor for a pendulum in an interferometric
gravitational waves detector”. In: \emph{Physics Letters A} 272.1 (2000),
pp. 39–45. \textsc{issn}: 0375-9601. \textsc{doi}: \href
{https://doi.org/https://doi.org/10.1016/S0375-9601(00)00411-4} {\nolinkurl
{https://doi.org/10.1016/S0375-9601(00)00411-4}}. \textsc{url}: \url
{https://www.sciencedirect.com/science/article/pii/S0375960100004114}.
{}
\bibitem{Hopcroft-silicon}
Matthew A. Hopcroft, William D. Nix, and Thomas W. Kenny. “What is the
Young’s Modulus of Silicon?” In: \emph{Journal of Microelectromechanical
Systems} 19.2 (2010), pp. 229–238. \textsc{doi}: \href
{https://doi.org/10.1109/JMEMS.2009.2039697} {\nolinkurl
{10.1109/JMEMS.2009.2039697}}.
{}
\bibitem{Mark-DD}
Mark Barton. \emph{Dissipation Dilution}. Technical Report LIGO-T070101-00.
Available from \url { https://dcc.ligo.org/LIGO-T070101/public }. 2008.
{}
\bibitem{nelson-modes}
Nelson L. Leon and Leonardo Gonz\'alez L\'opez. \emph{Analysis of the resonant modes
present in a novel cryogenic test-mass suspension for the Einstein Telescope
Gravitational Wave Observatory}. Technical Report ET-0045A-25. Available from
\url { https://apps.et-gw.eu/tds/ql/?c=17772 }. 2025.
{}
\bibitem{Joris-PhD-thesis}
J.V. van Heijningen. “Turn up the bass!: Low-frequency performance
improvement of seismic attenuation systems and vibration sensors for next
generation gravitational wave detectors”. English. PhD-Thesis - Research and
graduation internal. Vrije Universiteit Amsterdam, 2018. \textsc{url}: \url
{https://gwdoc.icrr.u-tokyo.ac.jp/cgi-bin/DocDB/ShowDocument?docid=7985}.
{}
\bibitem{Naticchioni-2018}
L. Naticchioni and on behalf of the Virgo Collaboration. “The payloads of
Advanced Virgo: current status and upgrades”. In: \emph{Journal of Physics:
Conference Series} 957.1 (2018), p. 012002. \textsc{doi}: \href
{https://doi.org/10.1088/1742-6596/957/1/012002} {\nolinkurl
{10.1088/1742-6596/957/1/012002}}. \textsc{url}: \url
{https://dx.doi.org/10.1088/1742-6596/957/1/012002}.
{}
\bibitem{Ushiba-2021}
Takafumi Ushiba et al. “Cryogenic suspension design for a kilometer-scale
gravitational-wave detector”. In: \emph{Classical and Quantum Gravity} 38.8
(2021), p. 085013. \textsc{doi}: \href
{https://doi.org/10.1088/1361-6382/abe9f3} {\nolinkurl
{10.1088/1361-6382/abe9f3}}. \textsc{url}: \url
{https://dx.doi.org/10.1088/1361-6382/abe9f3}.
{}
\bibitem{Bersanetti-PhD-thesis}
Diego Bersanetti. “Development of a New Lock Acquisition Strategy for the Arm
Cavities of Advanced Virgo”. English. PhD Thesis. University of Genova, 2016.
\textsc{url}: \url {https://tds.virgo-gw.eu/ql/?c=13950}.
{}
\bibitem{Beker-PhD-thesis}
M.G. Beker. “Low-frequency sensitivity of next generation gravitational wave
detectors”. PhD thesis. Vrije Universiteit Amsterdam, 2013. \textsc{url}: \url
{https://hdl.handle.net/1871/41484}.
{}
\bibitem{DiPace2020}
Sibilla Di Pace et al. “Thermal noise study of a radiation pressure noise limited
optical cavity with fused silica mirror suspensions”. In: \emph{The European
Physical Journal D} 74.11 (2020), p. 227. \textsc{issn}: 1434-6079. \textsc{doi}:
\href {https://doi.org/10.1140/epjd/e2020-10183-7} {\nolinkurl
{10.1140/epjd/e2020-10183-7}}. \textsc{url}: \url
{https://doi.org/10.1140/epjd/e2020-10183-7}.
{}
\bibitem{selleri-report}
Stefano Selleri. \emph{Optical Cavities as Springs – Antisprings}. Technical
Report ET-0062A-25. Available from \url {
https://apps.et-gw.eu/tds/ql/?c=17790 }. 2025.
{}
\bibitem{leonardo-report}
Leonardo Gonz\'alez L\'opez. \emph{Study of silicon spring blades for a novel
cryogenic test-mass suspension system for the Einstein Telescope Gravitational
Wave Observatory}. Technical Report ET-0043A-25. Available from \url {
https://apps.et-gw.eu/tds/ql/?c=17770 }. 2025.
{}
\bibitem{okhotin-si}
A.S. Okhotin et al. \emph{Thermophysical Properties of Semiconductors}. Cited
by [35]. Moscow: Atom Publishing House, 1972.
{}
\bibitem{PhysRevD.108.123009}
Xhesika Koroveshi et al. “Cryogenic payloads for the Einstein Telescope: Baseline
design with heat extraction, suspension thermal noise modeling, and sensitivity
analyses”. In: \emph{Phys. Rev. D} 108 (12 2023), p. 123009. \textsc{doi}: \href
{https://doi.org/10.1103/PhysRevD.108.123009} {\nolinkurl
{10.1103/PhysRevD.108.123009}}. \textsc{url}: \url
{https://link.aps.org/doi/10.1103/PhysRevD.108.123009}.
{}
\bibitem{PhysRevD.97.102001}
Kentaro Komori et al. “Direct approach for the fluctuation-dissipation theorem
under nonequilibrium steady-state conditions”. In: \emph{Phys. Rev. D} 97 (10
2018), p. 102001. \textsc{doi}: \href
{https://doi.org/10.1103/PhysRevD.97.102001} {\nolinkurl
{10.1103/PhysRevD.97.102001}}. \textsc{url}: \url
{https://link.aps.org/doi/10.1103/PhysRevD.97.102001}.
{}
\bibitem{app12178827}
Florian Amann et al. “Tunnel Configurations and Seismic Isolation Optimization
in Underground Gravitational Wave Detectors”. In: \emph{Applied Sciences}
12.17 (2022). \textsc{issn}: 2076-3417. \textsc{doi}: \href
{https://doi.org/10.3390/app12178827} {\nolinkurl {10.3390/app12178827}}.
\textsc{url}: \url {https://www.mdpi.com/2076-3417/12/17/8827}.
{}
\bibitem{si-properties-russia}
\emph{Silicon thermal properties}. \url
{https://www.ioffe.ru/SVA/NSM/Semicond/Si/thermal.html}. Accesed on
27-09-2024.
\end{thebibliography}
\end{document}